\documentclass[a4paper,11pt]{article}

\usepackage{jheppub}
\usepackage{bm}
\usepackage{amsmath,amssymb,bm}
\usepackage{graphicx}
\newcommand{\be}{\begin{equation}}
\newcommand{\ee}{\end{equation}}
\newcommand{\bea}{\begin{eqnarray}}
\newcommand{\eea}{\end{eqnarray}}

\newcommand{\eq}[1]{Eq.~(\ref{eq:#1})}

\newcommand{\del}{\partial}

\newcommand{\bra}{\langle}
\newcommand{\ket}{\rangle}
\newcommand{\calO}{{\cal O}}

\newcommand{\eg}{{\it e.g.}}

\bmdefine{\bmq}{{\bm{q}}}
\bmdefine{\bmk}{{\bm{k}}}
\bmdefine{\bmx}{{\bm{x}}}
\bmdefine{\bmy}{{\bm{y}}}
\bmdefine{\bmr}{{\bm{r}}}
\bmdefine{\bmnabla}{{\bm{\nabla}}}
\bmdefine{\bmA}{ \bm{A} }
\bmdefine{\bmD}{ \bm{D} }

\bmdefine{\bmPhi}{ \bm{\Phi} }
\bmdefine{\bmPsi}{ \bm{\Psi} }
\bmdefine{\bmcalO}{ \bm{\mathcal{O}} }

\newcommand{\calL}{{\cal L}}
\newcommand{\calM}{{\cal M}}

\newcommand{\vecx}{\vec{x}}

\newcommand{\nq}{\mathfrak{q}}
\newcommand{\nQ}{\mathfrak{Q}}
\newcommand{\nw}{\mathfrak{w}}

\bmdefine{\bmg}{{\bm{g}}}
\bmdefine{\bmR}{{\bm{R}}}

\newcommand{\mfh}{\mathfrak{h}}
\newcommand{\mfw}{\nw}
\newcommand{\mfq}{\nq}
\newcommand{\mfQ}{\nQ}
\newcommand{\mfA}{\mathfrak{A}}

\newcommand{\dw}{\delta\nw}
\newcommand{\dq}{\delta \nq}
\newcommand{\dqs}{\delta(\nq^2)}
\newcommand{\deta}{\delta\eta}

\title{Nonuniqueness of Green's functions at special points}

\author[a,1]{Makoto Natsuume,%
\note{Also at Department of Particle and Nuclear Physics, 
SOKENDAI (The Graduate University for Advanced Studies), 1-1 Oho, 
Tsukuba, Ibaraki, 305-0801, Japan;
 Department of Physics Engineering, Mie University, 
 Tsu, 514-8507, Japan.}
}
\author[b]{Takashi Okamura}

\affiliation[a]{KEK Theory Center, \\
Institute of Particle and Nuclear Studies, \\
High Energy Accelerator Research Organization,\\
Tsukuba, Ibaraki, 305-0801, Japan}
\affiliation[b]{Department of Physics, \\
Kwansei Gakuin University, \\
Sanda, Hyogo, 669-1337, Japan}

\emailAdd{makoto.natsuume@kek.jp}
\emailAdd{tokamura@kwansei.ac.jp}

\arxivnumber{1905.12015}
\preprint{KEK-TH-2129} 
\abstract{
We investigate a new property of retarded Green's functions using AdS/CFT. The Green's functions are not unique at special points in complex momentum space. This arises because there is no unique incoming mode at the horizon and is similar to the ``pole-skipping" phenomenon in holographic chaos. Our examples include the bulk scalar field, the bulk Maxwell vector and scalar modes, and the shear mode of gravitational perturbations. In these examples, the special points are always located at  $\omega_\star = -i(2\pi T)$ with appropriate values of complex wave number.}


\begin{document}

%\begin{flushright}
%        \today
%        {\it Preliminary version}
%\end{flushright}
\maketitle
\flushbottom

%%%%%%%%%
\section{Introduction and Summary}%\label{sec:}
%%%%%%%%%

The retarded Green's function is a very important object in physics and its property has been extensively studied. The AdS/CFT duality \cite{Maldacena:1997re,Witten:1998qj,Witten:1998zw,Gubser:1998bc} allows us to compute the Green's function at strong coupling and to provide us more information (see, \eg, Refs.~\cite{CasalderreySolana:2011us,Natsuume:2014sfa,Ammon:2015wua,Zaanen:2015oix,Hartnoll:2016apf}). The purpose of this paper is to explore a new aspect of the Green's function using AdS/CFT%
\footnote{While this paper and the companion paper \cite{Natsuume:2019sfp} are in preparation, there appeared preprints \cite{Grozdanov:2019uhi,Blake:2019otz} which have some overlap with ours. }. 

Our work is motivated from the recent development of holographic chaos. The holographic chaos has been studied using out-of-time-ordered correlation functions $C$ \cite{Shenker:2013pqa,Roberts:2014isa,Shenker:2014cwa,Maldacena:2015waa}. However, recently it is claimed that the chaotic behavior can be seen even at the level of retarded Green's functions. This phenomenon is known as ``pole-skipping" \cite{Grozdanov:2017ajz,Blake:2018leo}. The pole-skipping claims that retarded Green's function is not unique at a ``special point" in momentum space. From the location of the special point, one can extract the Lyapunov exponent $\lambda$ and the butterfly velocity $v_B$. 

More explicitly, consider the energy-density correlators. $C$ behaves as
\begin{align}
C(t,x) \simeq e^{\lambda(t-x/v_B)} = e^{-i\omega_\star t+iq_\star x}~.
%\label{eq:}
%
\end{align}
The pole-skipping claims that $(\lambda,v_B)$ or $(\omega_\star,q_\star)$ are determined from the retarded Green's function. Note that $\omega_\star$ is located in the upper-half $\omega$-plane. Generically, one would write the function as
\begin{align}
G^R_{T^{00}T^{00}} (\omega,q) = \frac{b(\omega,q)}{a(\omega,q)}~.
\label{eq:2pt}
\end{align}
The pole-skipping claims that 
\begin{align}
a(\omega_\star,q_\star) = b(\omega_\star,q_\star) = 0~.
%\label{eq:}
%
\end{align}
Then, naively $G^R=0/0$, but more precisely, $G^R$ is not uniquely determined at the special point. Near the special point, 
\begin{align}
G^R = \frac{ \delta\omega (\del_\omega b)_\star + \delta q (\del_q b)_\star +\cdots }{\delta\omega (\del_\omega a)_\star + \delta q (\del_q a)_\star +\cdots }
= \frac{ (\del_\omega b)_\star + \frac{\delta q}{\delta\omega} (\del_q b)_\star +\cdots }{ (\del_\omega a)_\star + \frac{\delta q}{\delta\omega} (\del_q a)_\star +\cdots }~.
%\label{eq:}
%
\end{align}
Then, the Green's function at the special point is not unique because it depends on the slope $\delta q/\delta\omega$. Also, it is argued that the ``would-be" pole $a(\omega_\star,q_\star)=0$ is related to a hydrodynamic pole. Namely, in hydrodynamic limit where $\omega, q \to 0$, $a(\omega,q)=0$ at the sound pole. If the sound pole is analytically continued to a pure imaginary $q$ and if it is extended to finite $q$, it coincides with the special point. 

The pole-skipping and the special point have been discussed for the energy-density Green's functions. A natural question is whether there are similar phenomena for the other Green's functions. This is the question we address in this paper. 

It turns out that many Green's functions possess special points in complex $(\omega,q)$-plane in the sense that Green's functions is not unique. 
The nonuniqueness of Green's function is in common with holographic chaos, but there are differences:
\begin{itemize}
\item 
The special points are located in the lower-half $\omega$-plane in our examples, so they do not seem to indicate chaotic behaviors. The special points are always located at $\omega_\star=-i(2\pi T)$.
\item
The special points are not necessarily related to hydrodynamic poles in our examples. There are special points for systems which lack a hydrodynamic pole (\eg, bulk scalar field).
\end{itemize}
In Ref.~\cite{Natsuume:2019sfp}, the special point of the sound mode is characterized as follows:
\begin{quote}
The bulk field equation has a regular singularity at the horizon $r=1$, but at the special point, it becomes a regular point in the incoming Eddington-Finkelstein (EF) coordinates.
\end{quote}
We use this criterion to explore special points of various Green's functions. Our examples include the bulk scalar field, the bulk Maxwell vector and scalar modes, and the shear mode of gravitational perturbations. 

In these examples, the field equation typically takes the following form near the horizon $r=1$%
\footnote{We assume that the horizon is nondegenerate. When the horizon is degenerate, the field equation has an irregular singularity.}:
\begin{align}
\phi'' + \frac{1-i\nw}{r-1}\phi' + \frac{P(\nw,\nq)}{r-1}\phi \sim 0~, \quad (r\to1)
%\label{eq:}
%
\end{align}
where
\begin{align}
\nw= \frac{\omega}{2\pi T}~, \quad \nq = \frac{q}{2\pi T}~,
%\label{eq:}
%
\end{align}
and $P$ is some function.

In general, the field equation has a regular singularity at the horizon. The equation has 2 solutions, an incoming mode and an outgoing mode. Since we are interested in the retarded Green's function, we select an incoming mode. However, by choosing $i\nw=1$ and by appropriately choosing $\nq$, one may eliminate the singularity in front of $\phi'$ and $\phi$ terms. Then, the point $r=1$ becomes a regular point. As a result, two solutions become regular there and are written as Taylor series. This is our basic observation. Note that the above criterion is a sufficient condition for the existence of two regular solutions but not the necessary condition.

At the special point, there are two regular solutions. Actually, as we see in details, the incoming-wave boundary condition is not uniquely defined at the special point and depends on the slope $\delta q/\delta\omega$, and both solutions must be included. Namely, there is no unique incoming mode at the special point. As a result, the Green's function is not unique. 

We close this section with a list of special points:
%\begin{alignat}{2}
%
%&\text{scalar: }&& A_v~, A_x~, A_r~, \\
%&\text{vector: }&& A_y, A_z~.
%\label{eq:}
%
%\end{alignat}
\begin{subequations}
%\label{eq:}
\begin{alignat}{4}
&\text{Massive scalar (SAdS$_{p+2}$): } && \nw_\star = -i~, \quad &&\nq_\star^2 = -\frac{2p(p+1)+4m^2}{(p+1)^2}~. \\
&\text{Maxwell vector (SAdS$_{p+2}$): } && \nw_\star = -i~, &&\nq_\star^2 = -\frac{2(p-2)}{p+1}~. \\
&\text{Maxwell scalar (SAdS$_{p+2}$): } && \nw_\star = -i~, &&\nq_\star^2 =  +\frac{2(p-2)}{p+1}~. \\
&\text{Gravitational tensor (SAdS$_{p+2}$): } && \nw_\star = -i~, \quad &&\nq_\star^2 = -\frac{2p}{p+1}~. \\
&\text{Gravitational shear (SAdS$_4$): } && \nw_\star = -i~, &&\nq_\star^2 = +\frac{4}{3}~. \\
&\text{Gravitational sound (SAdS$_4$): } && \nw_\star = +i~, &&\nq_\star^2 = -\frac{4}{3}~.
%\label{eq:}
%
\end{alignat}
\end{subequations}

%%%%%%%%%
\section{Preliminaries}%\label{sec:}
%%%%%%%%%

For illustration, we mostly consider the Schwarzschild-AdS$_{p+2}$ (SAdS$_{p+2}$) black hole background%
\footnote{We use upper-case Latin indices $M, N, \ldots$ for the $(p+2)$-dimensional bulk spacetime coordinates and use Greek indices $\mu, \nu, \ldots$ for the $(p+1)$-dimensional boundary coordinates. The boundary coordinates are written as  $x^\mu = (t, x^i) =(t, \vecx)=(t,x,y,\cdots)$. }%Lower-case Latin indices $a,b,\cdots$ are used for the 2-dimensional subspace $(v,r)$.}
:
\begin{align}
ds^2 &= r^2(-fdt^2+d\vecx_p^2) + \frac{dr^2}{r^2f}~,\\
f &= 1-r^{-p-1}~.
%\label{eq:}
%
\end{align}
For simplicity, we set the AdS radius $L=1$ and the horizon radius $r_0=1$. The Hawking temperature is given by $2\pi T=(p+1)/2$. Below we work with the incoming Eddington-Finkelstein (EF) coordinates. Using the tortoise coordinate $dr_*:=dr/(r^2f)$ and $v=t+r_*$, the metric becomes
\be
ds^2 = r^2(-fdv^2+d\vecx_p^2) +2dvdr~.
%\label{eq:}
%
\ee
We consider various perturbations in the black hole background. We often use the SAdS background, but the extension to the generic $f$ is straightforward. 

In this paper, we consider the scalar field, Maxwell field, and gravitational perturbations. We consider the perturbations of the form
\begin{align}
\phi(r) \, e^{-i\omega v +iqx}~.
%\label{eq:}
%
\end{align}
% v2
The perturbations are decomposed under the transformation of boundary spatial coordinate $x^i$. 
The Maxwell perturbations $A_M$ are decomposed as
\begin{align}
\text{scalar mode (diffusive mode): }& A_v~, A_x~, A_r~, \\
\text{vector mode: }& A_y~.
%\label{eq:}
%
\end{align}
%\begin{alignat}{2}
%
%&\text{scalar: }&& A_v~, A_x~, A_r~, \\
%&\text{vector: }&& A_y, A_z~.
%\label{eq:}
%
%\end{alignat}
For example, the scalar mode transforms as scalar under the transformation. 
Similarly, for $p=2$, gravitational perturbations are decomposed as 
\begin{align}
\text{scalar mode (sound mode): }& h_{vv}~, h_{vr}~, h_{rr}~, h_{vx}~, h_{rx}~, h_{xx}~, h_{yy}~, \\
\text{vector mode (shear mode): }& h_{vy}~, h_{ry}, h_{xy}~.
%\label{eq:}
%
\end{align}
In general, one also has a tensor mode, but for $p=2$, there is no tensor mode. The sound mode is analyzed previously \cite{Grozdanov:2017ajz,Blake:2018leo,Natsuume:2019sfp}, and we analyze the shear mode. The diffusive, sound, and shear modes have hydrodynamic poles. On the other hand, the Maxwell vector mode and a generic scalar field do not have a hydrodynamic pole. 

Normally, one fixes the gauge $A_r=h_{rM}=0$, and one constructs gauge-invariant variables which are invariant under the residual gauge transformation. This is the formalism advocated \eg, by Kovtun and Starients \cite{Kovtun:2005ev}. Instead, we do not fix the gauge and carry out analysis in a fully gauge-invariant manner. This is the formalism developed by Kodama and Ishibashi \cite{Kodama:2003jz}.

%%%%%%%%%
\section{Scalar field}%\label{sec:}
%%%%%%%%%

%%------------------
\subsection{SAdS background}%\label{sec:}
%%------------------

To illustrate a special point, consider a real scalar field. The field equation $(\nabla^2-m^2)\phi=0$ becomes 
\begin{align}
\phi'' + \left( \frac{f'}{f} + \frac{p+2}{r} - \frac{2i\omega}{r^2f} \right) \phi' - \frac{m^2r^2+pi\omega r+q^2}{r^4 f} \phi = 0~.
\label{eq:EOM_scalar}
\end{align}
For the SAdS$_{p+2}$ background, the field equation near the horizon behaves as
\begin{align}
\phi'' + \frac{ 1-\frac{2i\omega}{p+1} }{ r-1 }\phi' - \frac{m^2+pi\omega+q^2}{(p+1)(r-1)}\phi \sim 0~,
\quad(r\to1)~.
\label{eq:scalar_horizon}
\end{align}
For a generic $(\nw,\nq)$, the equation has a regular singularity at $r=1$. 

We consider the perturbation of the form $\phi(r) \, e^{-i\omega v +iqx}$. Since the horizon is a regular singularity, one can solve the equation by a power series expansion around $r=1$:
\begin{align}
\phi(r) = (r-1)^\lambda \sum_{n=0}\, a_n\, (r-1)^n~.
%\label{eq:}
%
\end{align}
At the lowest order, one gets the indicial equation $\lambda(\lambda-i\nw)=0$ and obtains
\begin{align}
\lambda_1=0~, \quad \lambda_2 = i\nw~.
%\label{eq:}
%
\end{align}
The coefficient $a_n$ is obtained by a recursion relation. The $\lambda_1$-mode represents the incoming mode since $e^{-i\omega v} = e^{-i\omega(t+r_*)}$. In the incoming EF coordinates, the incoming wave is a Taylor series. The $\lambda_2$-mode represents the outgoing mode since an outgoing mode is written as
\begin{align}
e^{-i\omega(t-r_*)} =e^{-i\omega v} e^{2i\omega r_*} \simeq e^{-i\omega v} (r-1)^{2i\omega/f'(1)} = e^{-i\omega v} (r-1)^{i\nw}~.
%\label{eq:}
%
\end{align}

For the SAdS$_{p+2}$ background, the regular singularity at $r=1$ becomes a regular point when
\be
\nw_\star=-i~, \quad\nq_\star^2 = -\frac{2p(p+1)+4m^2}{(p+1)^2}~.
\label{eq:sp_scalar}
\ee
We call it a ``special point." Then, one has 2 regular solutions:
\begin{align}
\lambda_1=0~, \quad \lambda_2 = 1~.
%\label{eq:}
%
\end{align}
At the special point, the field equation becomes
\begin{align}
\calL_\phi \phi_\star := \phi_\star'' + \left( \frac{f'}{f} + \frac{p+2}{r} - \frac{p+1}{r^2f} \right) \phi_\star' - \frac{2m^2(r+1)+p(p+1)}{2r^4 f}(r-1) \phi_\star = 0~.
%\label{eq:}
%
\end{align}
A few comments are in order:
\begin{itemize}
\item
While the indicial equation itself gives no constraint on $\nq$, it is necessary to eliminate the singularity in front of the $O(\phi)$ term in \eq{scalar_horizon}. When $i\nw=1$, $\lambda_1$ and $\lambda_2$ differ by an integer. In such a case, the smaller root fails to produce the independent solution since the recursion relation breaks down at some $a_n$. Instead, the second solution would contain a $\ln(r-1)$ term and is not regular at $r=1$. 

\item
Also, note that the $O(\phi)$ term is proportional to $(r-1)^{-1}$ not $(r-1)^{-2}$. In this sense, the $r=1$ singularity is a relatively ``mild" regular singularity. %This is an important property. 
If the point were a standard regular singularity, it is in general impossible to make it a regular point just by choosing $(\nw,\nq)$ since one has to eliminate $(r-1)^{-1}$ and $(r-1)^{-2}$ terms simultaneously%
\footnote{Even if one has a standard regular singularity $(r-1)^{-2}\phi$, one can always make it a mild regular singularity by setting $\phi=:(r-1)\varphi$. But the point is that one has a mild regular singularity in ``natural" variables: $\phi$ directly appears in physical quantities such as the energy-momentum tensor $T_{MN}$. This is related to a point in Ref.~\cite{Natsuume:2019sfp}. The paper uses natural variables to show regularity. }.
\end{itemize}

Write two solutions of $\phi$ as $\phi=C_1\phi_1+C_2\phi_2$. We consider linear perturbations, so an overall constant is not relevant, and the solution is parametrized by $C_2/C_1$. At the special point, the regular singularity becomes a regular point. As a result, $\phi_1$ and $\phi_2$ are both regular. Actually, the incoming mode is not uniquely determined. 

To see this, move away from the special point: 
\begin{align}
\nw=\nw_\star+\dw~, \nq=\nq_\star+\dq~, \phi=\phi_\star+\delta\phi~.
%\label{eq:}
%
\end{align}
Away from the special point, the field equation has a regular singularity at $r=1$ as usual, and the distinction between the incoming mode and the outgoing mode should be clear. Thus, we move away from the special point and approach the special point $\dw,\dq\to0$. 
In this way, one expects to obtain the incoming mode at the special point. However, as we see below, one cannot uniquely determine the incoming mode at the special point since the incoming mode depends on the slope $\dq/\dw$ how one approaches the special point.
 
The field equation is expanded as
\begin{align}
0&=\calL_\phi \delta\phi +j_\phi~,
\label{eq:EOM_scalar_sp} \\
j_\phi & = -\frac{p+1}{2r^4f} i \dw \left( 2r^2\del_r+ pr + \frac{p+1}{2}\frac{\dqs}{i\dw} \right) \phi_\star~.
%j_\phi & = -\frac{p+1}{2r^4f} i \dw \left\{ 2r^2\phi_\star' + \left( pr + \frac{\dq}{\dw}\sqrt{2p(p+1)+4m^2} \right)\phi_\star \right\}
%\label{eq:}
%
\end{align}
The perturbation $\delta\phi$ obeys an inhomogeneous differential equation, and the source $j_\phi$ is given by the special point solution $\phi_\star$. We are interested in the retarded Green's function, so we impose the incoming-wave boundary condition on the perturbation $\delta\phi$. An incoming wave is written as a Taylor series in the incoming EF coordinates, so the homogeneous part must be expanded as a Taylor series as well. However, the source contains $f^{-1}$ and is proportional to $(r-1)^{-1}$, so the equation in general produces an outgoing mode. To avoid this, we require that the source term is also written as a Taylor series. 

From \eq{EOM_scalar_sp}, $\delta\phi$ is regular at the horizon when the special point solution $\phi_\star$ satisfies
\begin{align}
\left. \frac{\phi_\star'}{\phi_\star} \right|_{r=1} = 
-\frac{p}{2} - \frac{p+1}{4}\frac{\dqs}{i\dw}~.
%-\frac{p}{2} - \frac{1}{2}\frac{\dq}{\dw} \sqrt{ 2p(p+1) + 4m^2 }
\label{eq:BC_scalar}
\end{align}
This is the incoming-wave boundary condition for $\phi_\star$. The boundary condition is not unique, or the boundary condition does not uniquely determine $\phi_\star$ and depends on $\dq/\dw$. Conversely, given a $\dq/\dw$, one has to choose the combination $C_2/C_1$ appropriately. 

Now, let us consider the Green's function at the special point. Asymptotically, $\phi$ behaves as
\begin{align}
\phi &\sim \phi^{(0)} r^{-\Delta_-} + \phi^{(1)} r^{-\Delta_+}~, \quad (r\to\infty) \\
\Delta_\pm &= \frac{p+1}{2} \pm \sqrt{ \left(\frac{p+1}{2}\right)^2 +m^2 }~.
\label{eq:asymp_scalar}
\end{align}
The falloffs depend on $C_2/C_1$ which is normally fixed by the incoming-wave boundary condition but is not fixed at the special point. Following the standard Hamilton-Jacobi procedure, one obtains the operator expectation value:
\begin{align}
\bra\calO\ket = \left. r^{\Delta_+-\Delta_--1}(-r^2\del_r+i\omega)(r^{\Delta_-}\phi) \right|_{r\to\infty}~.
%\label{eq:}
%
\end{align}
Here, we add the counterterm
\begin{align}
S_\text{CT} = -\frac{\Delta_-}{2}\int d^{p+1}x \sqrt{-\gamma}\phi^2~,
%\label{eq:}
%
\end{align}
where $\gamma_{\mu\nu}$ is the $(p+1)$-dimensional boundary metric $\gamma_{\mu\nu}dx^\mu dx^\nu = r^2(-fdt^2+d\vecx_p^2)$. One would need an additional counterterm of the form 
\begin{align}
S_\text{CT2} \propto \int d^{p+1}x \sqrt{-\gamma}\gamma^{\mu\nu}\del_\mu\phi\del_\nu\phi
\quad \text{for} \quad
m^2 \geq 1-\left(\frac{p+1}{2}\right)^2
\label{eq:CT2_scalar}
\end{align}
since the term becomes a relevant operator. 
The retarded Green's function is given by
\begin{align}
G_\calO
:= -\frac{\delta\bra\calO\ket}{\delta \phi^{(0)}} 
= -(\Delta_+-\Delta_-) \frac{ \phi^{(1)} }{ \phi^{(0)} }~.
%\label{eq:}
%
\end{align}
The Green's function is not unique at the special point since the falloffs depend on $C_2/C_1$ or depend on $\dq/\dw$.

As a solvable example, consider the $p=1$ case. The background metric is the BTZ black hole although in our case $x$ is not compact. 
The field equation at the special point becomes
\begin{align}
\phi_\star'' + \frac{3r+1}{r(r+1)}\phi_\star' - \frac{m^2r+m^2+1}{r^2(r+1)}\phi_\star = 0~.
%\label{eq:}
%
\end{align}
The horizon $r=1$ is clearly a regular point. The solution is given by
\begin{align}
\phi_\star = \frac{r}{1+r} (\phi_\star^{(0)} r^{-\Delta_-} + \phi_\star^{(1)} r^{-\Delta_+})~,
\label{eq:sol_scalar}
\end{align}
where $\Delta_\pm = 1\pm\sqrt{1+m^2}$. Imposing the boundary condition \eqref{eq:BC_scalar} at the special point $\nq_\star= i\sqrt{1+m^2}$, one obtains
\begin{align}
\frac{ \phi_\star^{(1)} }{ \phi_\star^{(0)} } = \frac{ 1+ \frac{\dq}{\dw} }{ 1- \frac{\dq}{\dw} }~.
%\label{eq:}
%
\end{align}
The retarded Green's function is given by
\begin{align}
G_\calO^\star 
%= \frac{\delta\calO}{\delta \phi^{(0)}} 
= -(\Delta_+-\Delta_-) \frac{ \phi^{(1)} }{ \phi^{(0)} }
= -(\Delta_+-\Delta_-) \frac{ 1+ \frac{\dq}{\dw} }{ 1- \frac{\dq}{\dw} }~.
%\label{eq:}
%
\end{align}

As is clear from this expression, the Green's function at the special point is not unique since it depends on $\dq/\dw$. By choosing the slope appropriately, one can have a pole or a zero. This structure is the same as the pole-skipping phenomenon in holographic chaos.

In this section, we consider a massive neutral scalar, but a few extension is possible:
\begin{itemize}
\item 
One may consider a massless scalar field. The gravitational tensor mode (for $p>2$) also falls in this class since it in general reduces to a minimally-coupled massless scalar field. The special point, the boundary condition, and the $p=1$ special point solution are given by setting $m=0$ in Eqs.~\eqref{eq:sp_scalar}, \eqref{eq:BC_scalar}, and \eqref{eq:sol_scalar}, respecitively. For example, the special point is located at
\be
\nw_\star=-i~, \quad\nq_\star^2 = -\frac{2p}{p+1}~.
%\label{eq:}
%
\ee
In order to compute the $p=1, m=0$ Green's function, one needs to take into account the counterterm \eqref{eq:CT2_scalar}.

\item
One may consider a complex scalar field. For example, consider an Einstein-Maxwell-complex scalar system. The system is hard to solve in general, but again consider a test scalar $\psi$ in a background metric and in a background Maxwell field $A_v$. In this case, $\nw_\star$ does not change in the gauge $A_v(r=1)=0$, but $\nq_\star$ changes and depends on the electric field on the horizon.
\end{itemize}

%%------------------
\subsection{Generic background}%\label{sec:}
%%------------------

The above discussion can be extended to an arbitrary background which takes the form
\begin{align}
ds^2 &= -F(r)dt^2 + \frac{dr^2}{F(r)} + R(r)^2 d\vecx_p^2 \\
& = -F(r) dv^2+2dvdr +\cdots~.
%F(r) &\sim F'(1) (r-1)~, \quad (r\sim1)
%\label{eq:}
%
\end{align}
This is the general static background with $SO(p)$ invariance. We assume $F(r) \sim F'(1)(r-1)$ near the horizon $r=1$. The Hawking temperature is given by $2\pi T =F'(1)/2$. The field equation is given by
\begin{align}
0 &= \phi'' + \frac{1}{F}\left( \frac{(FR^p)'}{R^p} - 2i\omega \right) \phi' 
- \frac{1}{F}\left( \frac{p i\omega R'}{R} + \frac{q^2}{R^2} +m^2 \right) \phi 
\label{eq:EOM_scalar2} \\
&\sim \phi''+\frac{1-i\nw}{r-1} \phi' 
- \left. \left( \frac{m^2}{F'} + \frac{ \nq^2 F'+2pi\nw R R' }{4R^2} \right) \right|_{r=1} \frac{\phi}{r-1}~, \quad (r\sim 1)~.
%- \left( \frac{m^2}{F'(1)} + \frac{ F'(1)\{\nq^2 F'(1)+2pi\nw R(1) R'(1)\}}{4R(1)^2} \right) \frac{\phi}{r-1} \quad (r\sim 1)
%\label{eq:}
%
\end{align}
The special point is located at
\begin{align}
\nw_\star=-i~, \quad \nq_\star^2 = - \left. \frac{R^2}{F'^2} \left( 2p F'\frac{R'}{R} + 4m^2 \right) \right|_{r=1}~.
%\label{eq:}
%
\end{align}
Expanding the field equation around the special point, one obtains the boundary condition at $r=1$ for special point solution:
\begin{align}
\left. \frac{\phi_\star'}{\phi_\star} \right|_{r=1} = 
\left. -\frac{p}{2}\frac{R'}{R} 
-\frac{F'}{4R^2}\frac{\dqs}{i\dw} \right|_{r=1}~.
%-\frac{1}{2R(1)}\frac{\dq}{\dw} \sqrt{ 2p F'(1)\frac{R'(1)}{R(1)} + 4m^2 }
\label{eq:BC_scalar2}
\end{align}

The existence of a special point is generic. 
%v1.1
It holds not only to asymptotically AdS black holes but also to the other black branes: for example, asymptotically flat black branes, de~Sitter black holes, the Rindler space, Lifshitz black holes, and hyperscaling-violating black holes. 
%It holds not only to asymptotically AdS black holes but also to asymptotically flat black branes, de~Sitter black holes, and the Rindler space. 
Unlike AdS black holes, the boundary interpretation is not fully understood for these spacetimes. But from the bulk point of view, the incoming mode is not uniquely defined at special points for these spacetimes as well. The existence of a special point and its location depend only on the near-horizon behavior and do not depend on the asymptotic structure.

%%%%%%%%%
\section{Maxwell field}%\label{sec:}
%%%%%%%%%

The gauge-invariant variables for the Maxwell field and gravitational perturbations are discussed in Ref.~\cite{Natsuume:2019sfp}. See App.~A in the paper for the details. 

%%------------------
\subsection{Maxwell vector mode}%\label{sec:}
%%------------------

The vector perturbation $A_y$ is gauge-invariant by itself. The Maxwell equation becomes
\begin{align}
&A_y'' + \left( \frac{f'}{f} + \frac{p}{r} - \frac{2i\omega}{r^2f} \right) A_y' - \frac{(p-2)i\omega r+q^2}{r^4f} A_y =0~.
\label{eq:eom_Ay}
\end{align}
For the SAdS$_{p+2}$ background, the special point is located at%
% v2
\footnote{
Note added in v2: For the $p=2$ Maxwell field, $\nq_\star=0$. According to Ref.~\cite{Blake:2019otz}, this is not a special point but an ``anomalous point." But it is not really necessary to distinguish anomalous points from the other special points. See Ref.~\cite{Natsuume:2019vcv} for the details.
}
\begin{align}
\nw_\star=-i~, \quad\nq_\star^2 = -\frac{2(p-2)}{p+1}~.
%\label{eq:}
%
\end{align}
The field equation at the special point becomes
\begin{align}
&\calL_y A_y^\star
:= A_y^\star{}'' + \left( \frac{f'}{f} + \frac{p}{r} - \frac{p+1}{r^2f} \right) A_y^\star{}' - \frac{(p+1)(p-2)}{2r^4f} (r-1)A_y^\star =0~.
%\label{eq:eomr}
%
\end{align}
and the horizon is a regular point.
 
Just like the scalar field, expand the field equation near the special point:
\begin{subequations}
%\label{eq:}
\begin{align}
0&=\calL_y \delta A_y +j_y~, \\
j_y& = -\frac{p+1}{4r^4f}i\dw \left\{ 4r^2 \del_r + 2(p-2)(r-1)+(p+1)\frac{\deta}{i\dw} \right\} A_y^\star~, \\
\eta &:= \frac{\nq^2}{i\nw}~.
\label{eq:EOM_Ay_sp}
\end{align}
\end{subequations}
%where $\eta:=\nq^2/(i\nw)$.
The perturbative solution is regular at the horizon when the special point solution $A_y^\star$ satisfies
\begin{align}
\left. \frac{A_y^\star{}'}{A_y^\star} \right|_{r=1} = -\frac{p+1}{4} \frac{\deta}{i\dw}~.
\label{eq:BC_Ay}
\end{align}
Thus, $A_y^\star$ is not unique and depends on the slope $\deta/\dw$. Given the slope, one has to choose the ratio of 2 integration constants of $A_y^\star$ appropriately.

For the $p=2$ case, one can obtain the special point solution, so one can obtain the Green's function at the special point explicitly. The special point solution is given by
% C2 sign changed
\begin{align}
A_y^\star &= C_1 - \frac{C_2}{3} 
\left[
\frac{ (r-1) \exp\left\{ \sqrt{3}~\left( \arctan\frac{ 2r + 1 }{\sqrt{3}} - \frac{\pi}{2} \right)\right\} }
{ (r^2+r+1)^{1/2} }
-1
\right]
\label{eq:sol_vector} \\
&\sim C_1 + \frac{C_2}{r} + \cdots~, \quad (r\to\infty)~.
\end{align}
Imposing the boundary condition \eqref{eq:BC_Ay}, one obtains
\begin{align}
%
%\left. \frac{A_y^\star{}'}{A_y^\star} \right|_{r=1} &= \frac{ C_2 e^{-\frac{\pi}{2\sqrt{3}}} }{ \sqrt{3}(3C_1-C_2) } 
%\\
\frac{C_2}{C_1} &= \frac{-27\gamma}{ 9\gamma - 4\sqrt{3}\,e^{-\frac{\pi}{2\sqrt{3}}} } :=c_T^\star~,
%\label{eq:}
%
\end{align}
where $\gamma:=\deta/(i\dw)$. 

Again following the standard procedure, one obtains 
\begin{align}
\bra J^\mu \ket & = \left.  -\sqrt{-g} F^{r\mu} \right|_{r\to\infty}~,
%\Rightarrow
%& \left. \bra J^v \ket = -r^p F^{rv} = r^p F_{rv} \right|_{r\to\infty}~.\\
%& \left. \bra J^y \ket = -r^p F^{ry} = - r^{p-2}(r^2f F_{ry}+F_{vy}) \right|_{r\to\infty}~.
\label{eq:VEV_Maxwell}
\end{align}
where no counterterm is necessary for $p<3$. From the asymptotic behavior, one obtains
\begin{align}
\bra J^y \ket 
&= C_2 + \frac{3}{2}i\nw C_1~.
%\label{eq:}
%
\end{align}
Thus, the Green's function $\Pi_T^\star$ is given by
\begin{align}
\Pi_T^\star = - \frac{C_2}{C_1} - \frac{3}{2}i\nw = - c_T^\star -\frac{3}{2}~.
%\label{eq:}
%
\end{align}
The Green's function is often given by the ratio $C_2/C_1$, but this may not be true when one uses the EF coordinates.
 
%%------------------
\subsection{Maxwell scalar mode}%\label{sec:}
%%------------------

The gauge-invariant variables for the scalar mode are given by 
\begin{align}
\mfA_v &= A_v + \frac{\omega}{q}A_x~, 
\label{eq:inv_At} \\
\mfA_r &= A_r - \frac{1}{iq}A_x'~.
\label{eq:inv_Ar}
\end{align}
The Maxwell equation becomes
\begin{align}
0 &= \mfA_v' + \frac{q^2}{i\omega r^2} \mfA_v-\frac{1}{i\omega}\left( \omega^2-q^2 f \right)\mfA_r~,
\label{eq:eomv} \\
0 &= \mfA_r' + \left( \frac{f'}{f} + \frac{p}{r} - \frac{2i\omega}{r^2f} - \frac{q^2}{i\omega r^2} \right) \mfA_r +\frac{1}{f} \left( -\frac{q^2}{i\omega r^4} + \frac{p-2}{r^3}\right)\mfA_v~.
\label{eq:eomr}
\end{align}
For the SAdS$_{p+2}$ background, the special point is located at
\begin{align}
\nw_\star=-i~, \quad\nq_\star^2 = \frac{2(p-2)}{p+1}~.
%\label{eq:}
%
\end{align}
Note $\nq_\star$ is real. Recalling that the special point represents a ``would-be" pole, the pole is located in the physical region (lower-half $\omega$-plane and real $q$) and is interesting. Its implication is not clear to us though. (The special point of the gravitational shear mode also has this property. See next section.)

Expand the field equation near the special point:
\begin{subequations}
\label{eq:EOM_diffusive_sp}
\begin{align}
0 &= \delta\mfA_v' + \frac{p-2}{r^2} \delta\mfA_v + \left\{ (p-2)f+\frac{p+1}{2} \right\}\delta\mfA_r +j_v~,
%\label{eq:eomv} 
\\
0 &= \delta\mfA_r' + \left\{ \frac{f'}{f} - \frac{p+1}{r^2f} + \frac{2+p(r-1)}{r^2} \right\} \delta\mfA_r +\frac{(p-2)(r-1)}{r^4f}\delta\mfA_v +j_r~,
%\label{eq:eomr}
\displaybreak \\
j_v &= \frac{p+1}{2}i\dw \left\{ \frac{\deta}{i\dw} \frac{\mfA_v^\star}{r^2} + \left(1+\frac{\deta}{i\dw}f \right)\mfA_r^\star \right\}~, \\
j_r &= -\frac{p+1}{2f}i\dw \left\{ \frac{\deta}{i\dw} \frac{\mfA_v^\star}{r^4} + \left(2+\frac{\deta}{i\dw}f \right)\frac{\mfA_r^\star}{r^2} \right\}~.
\label{eq:sourcer}
\end{align}
\end{subequations}
The source $j_r$ contains $f^{-1}$. Thus, the perturbative solution is regular at the horizon when the special point solution satisfies
\begin{align}
\left. \frac{\mfA_r^\star}{\mfA_v^\star} \right|_{r=1} = -\frac{1}{2} \frac{\deta}{i\dw}~.
\label{eq:BC_diffusive}
\end{align}

Again consider the $p=2$ case. The special point solution $\mfA_v^\star$ is the same as the $A_y$ solution \eqref{eq:sol_vector}.
Imposing the boundary condition \eqref{eq:BC_diffusive}, one obtains
\begin{align}
\frac{C_2}{C_1} &= \frac{-27\gamma}{ 9\gamma + 4\sqrt{3}\,e^{-\frac{\pi}{2\sqrt{3}}} } :=c_L^\star~.
%\label{eq:}
%
\end{align}
Using \eq{VEV_Maxwell}, one obtains
\begin{align}
\bra J^v \ket 
&= \frac{\nq^2}{\nw^2-\nq^2} \left. \left( -r^2 \mfA_v' + i\omega \mfA_v \right) \right|_{r\to\infty} \\
&= \frac{\nq^2}{\nw^2-\nq^2} (C_2 + \frac{3}{2}i\nw C_1)~,
%\label{eq:}
%
\end{align}
where \eq{inv_At} is used in the first line. Thus, the Green's function $G_{vv}^\star$ is given by
% v2
\begin{align}
G_{vv}^\star &=: \frac{\nq^2}{\nw^2-\nq^2} \Pi_L^\star~,\\
\Pi_L^\star &= - \frac{C_2}{C_1} - \frac{3}{2}i\nw
= - c_L^\star - \frac{3}{2}~.
%\label{eq:}
%
\end{align}
% v1.3
$\Pi_L^\star$ and $\Pi_T^\star$ are not uniquely determined, but the product is constant%
\footnote{Note added in v2: This is a consequence of the self-duality of the bulk four-dimensional Maxwell field \cite{Herzog:2007ij,Natsuume:2019vcv}. }:
\begin{align}
\Pi_T^\star\Pi_L^\star = (2\pi T)^2~.
%\label{eq:}
%
\end{align}

One may worry about the $p=2$ special point $\nq_\star=0$. When $q=0$, various expressions become singular, and one can show that field equations become trivial. This is not really a problem however. We define the Green's function at the special point by taking the limit $\dw, \dq\to0$. What is really meaningful is the $q\neq0$ expressions such as \eq{EOM_diffusive_sp}.

The situation is somewhat similar to standard Green's functions in the hydrodynamic limit. Ref.~\cite{Kovtun:2005ev} computes the hydrodynamic limit of various Green's functions for the $\mathcal{N}=4$ SYM (or the SAdS$_5$ background). The Green's functions for the Maxwell scalar mode ($\Pi_L$), the shear mode ($G_\text{shear}$), and the sound mode ($G_\text{sound}$) behave as
\begin{subequations}
\begin{align}
  & \Pi_L
  = \frac{N_c^2\, T^2}{8}\, \frac{ \mfw^2 - \mfq^2 }{ i\, \mfw - \mfq^2 }
%  = - \frac{N_c^2\, T^2}{8}\, \frac{ i\, \mfw + \eta }{ 1 - \eta }
%  ~\xrightarrow{ |\, \mfw\, | \ll |\, \eta\, | }\,
%    - \frac{N_c^2\, T^2}{8}\, \frac{\eta}{ 1 - \eta }
  ~,
\label{eq:R_current-Green_L} \\
%  & \Pi^T
%  = - \frac{N_c^2\, T^2}{8}\, i\, \mfw
%  ~,
%\label{eq:R_current-Green_T} \\
% correcred
  & G_\text{shear}
  = \frac{\pi^2\, N_c^2\, T^4}{2}\, \frac{ \mfw^2 - \mfq^2 }{ i\, \mfw - \mfq^2/2 }
%  = - \pi^2\, N_c^2\, T^4\, \frac{ i\, \mfw + \eta }{ 1 - \eta/2 }
%  ~\xrightarrow{ |\, \mfw\, | \ll |\, \eta\, | }\,
%    - \pi^2\, N_c^2\, T^4\, \frac{\eta}{ 1 - \eta/2 }
  ~,
\label{eq:shear-Green} \\
  & G_\text{sound}
  = \pi^2\, N_c^2\, T^4\, \frac{ \mfq^2 }{ 3\, \mfw^2 - \mfq^2 }
%  = - \pi^2\, N_c^2\, T^4\, \frac{\eta}{ \eta + 3\, i\, \mfw }
%  = - \pi^2\, N_c^2\, T^4\,
%    \frac{ \frac{\eta}{i\, \mfw} }{ 3 + \frac{\eta}{i\, \mfw} }
%  = \frac{\pi^2\, N_c^2\, T^4}{2}\,
%  \frac{ 2\, \mfq^2 + i\, \mfw\, \left( 5\, \mfq^2 - 3\, \mfw^2 \right) }
%       { 3\, \mfw^2 - \mfq^2 + 2\, i\, \mfw\, \mfq^2 }
%  = - \pi^2\, N_c^2\, T^4\,
%  \frac{ 1 + i\, \mfw\, \left( 5 - 3\, \mfw^2/\mfq^2 \right)/2  }
%       { 1 - 3\, \mfw^2/\mfq^2 - 2\, i\, \mfw }
%  = - \frac{\pi^2\, N_c^2\, T^4}{2}\,
%  \frac{ i\, \mfw\, \left( 3\, i\, \mfw + \frac{5\, \mfq^2}{i\, \mfw} \right)
%    + \frac{\mfq^2}{i\, \mfw} }
%       { i\, \mfw\, \left( 3\, i\, \mfw - \frac{2\, \mfq^2}{i\, \mfw} \right)
%    + \frac{\mfq^2}{i\, \mfw} }
%  = - \frac{\pi^2\, N_c^2\, T^4}{2}\,
%  \frac{ 2\, \eta + i\, \mfw\, \left( 3\, i\, \mfw + 5\, \eta \right) }
%       { \eta + 3\, i\, \mfw\, \left( 1 - 2\, \eta/3 \right) }
  ~.
\label{eq:sound-Green}
\end{align}
\end{subequations}
The point $\nw=\nq=0$ is a ``special point" in the sense that the Green's functions depend on the slope $\nq^2/(i\nw)$. This is related to the fact that the bulk field equations become trivial when $\nw=\nq=0$.

%%------------------
\subsection{Special point as eigenvalue problem}%\label{sec:}
%%------------------

For the Maxwell scalar mode,  there are 2 variables, $(\mfA_v, \mfA_r)$. When there are more variables, it becomes tedious to carry out analysis in this way. 

One method is to use a ``master variable." Actually, the Maxwell scalar mode can be formulated in this manner. However, this method has some problems as well:
\begin{itemize}
\item First, it is often not easy to find a master variable. 
\item Second, a master variable may fail at particular points in momentum space including the special point. It is straightforward to use the full set of variables. The sound mode is an example \cite{Natsuume:2019sfp}.
\end{itemize}
Thus, we do not take this approach. Instead, we formulate the problem as an eigenvalue problem. 

One would write the field equation of the Maxwell scalar mode as 
\begin{align}
&0 = X'- M X~, \\
&X(r) := \begin{pmatrix} \mfA_v \\ \mfA_r \end{pmatrix}~,
\\
&M := \dfrac{f'(1)}{2}\,
  \begin{pmatrix} -\dfrac{\eta}{r^2} & -i \nw -\eta f \\
%         \dfrac{ \stackrel{\mathstrut}{1} }{f}\,
           \dfrac{1}{r^4f}\left( \eta -  \dfrac{ 2(p - 2)r }{ f'(1) } \right)~
       & \dfrac{\eta}{r^2}-\dfrac{2p}{rf'(1)}+\dfrac{2}{f}\left(\dfrac{i\nw}{r^2}-\dfrac{f'(r)}{f'(1)}\right) 
  \end{pmatrix}~.
%\label{eq:}
%
\end{align}
The horizon $r=1$ is a regular singularity, so the matrix $M$ can be expanded as
\begin{align}
M &= \frac{\calM}{r-1}+M_0+\cdots~, \\
\calM &:= 
  \begin{pmatrix} 0 & 0 \\
    \dfrac{\eta}{2} -  \dfrac{p - 2}{f'(1)}~ & i\nw-1
  \end{pmatrix}~.
%\label{eq:}
%
\end{align}
The solution can be written as a power series:
\begin{align}
X=(r-1)^\lambda \sum_{n=0}\, \chi_n\, (r-1)^n~.
%\label{eq:}
%
\end{align}
Substituting this into the field equation, at the lowest order, one obtains
\begin{align}
0 =(\lambda - \calM)\chi_0~.
%(\lambda+1 - \calM)\chi_1=m_0\chi_0~.
%\label{eq:}
%
\end{align}
This indicial equation is an eigenvalue equation for $\calM$. 
The horizon $r=1$ becomes a regular point when $\calM=0$, so
\begin{align}
\nw_\star=-i~, \quad \eta_\star = \frac{2(p-2)}{f'(1)} \quad\Rightarrow\quad \nq_\star^2 = \frac{2(p-2)}{f'(1)}~.
%\label{eq:}
%
\end{align}

The eigenvalues and the eigenvectors of $\calM$ are
\begin{align}
  & \lambda_1 = 0~,
& & \chi^{(1)}_0
  = \begin{pmatrix} 1 - i\nw \\
       \dfrac{\eta}{2} -  \dfrac{p - 2}{f'(1)}
    \end{pmatrix}
  ~, 
\\
  & \lambda_2 = i\nw -1~,
& & \chi^{(2)}_0
  = \begin{pmatrix} 0 \\ 1 \end{pmatrix}
  ~.
\end{align}
The mode with $\lambda_1$ ($\lambda_2$) corresponds to the incoming (outgoing) mode. However, at the special point, zero eigenvalues degenerate, and one cannot distinguish between the incoming and the outgoing modes%
\footnote{Two eigenvalues degenerate when $i\nw=1$. For a generic $\eta$ or $\nq^2$, eigenvectors are not independent, and $\chi^{(1)}_0 \propto \chi^{(2)}_0$. In such a case, there is only one independent power series solution, and the second solution would contain a $\ln(r-1)$ term. Since the log term is not regular, one can still distinguish between the incoming and the outgoing modes.}.

Near the special point, the distinction is clear, and $\chi^{(1)}_0$ is the incoming mode. Thus, expand the eigenvector near the special point:
$ \nw=\nw_\star+\dw~, \eta= \eta_\star+\delta\eta~. $
Then, $\chi^{(1)}_0$ becomes
\begin{align}
\chi^{(1)}_0  
 \sim \begin{pmatrix} - i\dw \\ \dfrac{\delta\eta}{2} \end{pmatrix}~.
%\label{eq:}
%
\end{align}
The eigenvector $\chi^{(1)}_0$ vanishes at the special point, so the eigenvector has the slope dependence near the special point. The slope dependence of the incoming-wave boundary condition comes directly from this property:
\begin{align}
\left. \frac{\mfA_r^\star}{\mfA_v^\star} \right|_{r=1} = -\frac{1}{2} \frac{\deta}{i\dw}~.
%\label{eq:BC_diffusive}
%
\end{align}

%%%%%%%%%
\section{Shear mode}%\label{sec:}
%%%%%%%%%

% based on special_pt_v5_0409.tex

For simplicity, we consider only the $p=2$ case. The gauge-invariant variables for the shear mode are 
\begin{align}
\mfh_{vy} &= h_{vy} +\frac{\omega}{q} h_{xy}~, \\
\mfh_{ry} &= h_{ry} -\frac{r^2}{iq}\left( \frac{ h_{xy} }{r^2} \right)'~.
%\label{eq:}
%
\end{align}
(See App.~A of Ref.~\cite{Natsuume:2019sfp}.) The field equations are given by
\begin{subequations}
\label{eq:EOM-mfh-vec-all}
\begin{align}
  & 0
  = \left( \frac{ \mfh_{vy} }{r^2} \right)'
  + M_{vv}(r)\, \frac{ \mfh_{vy} }{r^2}
  + M_{vr}(r)\, \mfh_{ry}
  ~,
\label{eq:EOM-hvy} \\
  & 0
  = \mfh_{ry}' + M_{rv}(r)\, \frac{ \mfh_{vy} }{r^2}
  + M_{rr}(r)\, \mfh_{ry}
  ~,
\label{eq:EOM-hry} \\
  & M_{vv}
%  := - \frac{2}{r}\,
%    \left( 1 - \frac{\mfQ^2 }{i\, \mfw}\, \frac{1}{r} \right)
  := \frac{2\, \mfQ^2}{i\, \mfw}\, \frac{1}{r^2}
  ~,
%\label{def-M_vv} \\
  \hspace{2.8truecm}
   M_{vr}
  := \frac{1}{r^2}\, \left( \frac{3}{2}\, i\, \mfw
  + \frac{2\, \mfQ^2}{i\, \mfw}\, f \right)
  ~,
\label{eq:def-Mvr} \\
  & M_{rv}
%%  := \frac{2}{r^3\, f}\,
%%    \left( 1 - \frac{\mfQ^2}{i\, \mfw}\, \frac{1}{r} \right)
%  := - \frac{ M_{vv} }{r^2\, f}\,
  := \frac{2}{r\, f}\,
    \left( 1 - \frac{\mfQ^2 }{i\, \mfw}\, \frac{1}{r} \right)
  ~,
%\label{def-M_rv} \\
  \hspace{1.3truecm}
   M_{rr}
  := \frac{3}{r\, f} \left\{ 1 - \frac{i\, \mfw}{r}
  - \left( 1 + \frac{2\, \mfQ^2}{i\, \mfw}\, \frac{1}{r} \right)\,
     \frac{f}{3} \right\}
  ~.
\label{eq:def-Mrr}
\end{align}
\end{subequations}
where $\nQ:=\sqrt{3}\nq/2$. For the SAdS$_4$ background, the special point is located at
\begin{align}
\nw_\star=-i~, \quad\nQ_\star^2 = 1~.
%\label{eq:}
%
\end{align}
For comparison, the special point of the sound mode is located at $\nw_\star=i, \nQ_\star^2=-1$.

The horizon becomes a regular point at the special point. Expand the field equation near the special point:
\begin{subequations}
\label{eq:exp-EOM-mfh-vec-all}
\begin{align}
   0
  &= \left( \frac{ \delta\mfh_{vy} }{r^2} \right)'
  + \frac{2}{r^2}\, \frac{ \delta\mfh_{vy} }{r^2}
  + \frac{1}{r^2}\, \left( \frac{7}{2} - \frac{2}{r^3} \right)\,
    \delta\mfh_{ry}
\nonumber \\
  &+ \frac{2\, \delta\eta}{r^2}\, \frac{ \mfh^\star_{vy} }{r^2}
  + \frac{3}{2}\, \frac{1}{r^2}\,
    \left( i\, \delta\mfw + 4\, \delta\eta\, \frac{f}{3} \right)\,
    \mfh^\star_{ry}
  ~,
\label{eq:exp-EOM-hvy} \\
   0
  &= \delta\mfh_{ry}'
  + \frac{2\, r}{1 + r + r^2}\, \frac{ \delta\mfh_{vy} }{r^2}
  + \frac{1}{r^2}\, \left( 2\, r - 5 + \frac{3}{1 + r + r^2}
    \right)\, \delta\mfh_{ry}
\nonumber \\
  &- \frac{2\, \delta\eta}{r^2\, f}\, \frac{ \mfh^\star_{vy} }{r^2}
  - \frac{3}{r^2\, f}\, \left( i\, \delta\mfw
  + 2\, \delta\eta\, \frac{f}{3} \right)\, \mfh^\star_{ry}
  ~.
\label{eq:exp-EOM-hry}
\end{align}
\end{subequations}
where $\eta:=\nQ^2/(i\nw)$. From the second line of \eq{exp-EOM-hry}, the perturbative solution is regular at the horizon when the special point solution satisfies
%
%\begin{subequations}
%\label{eq:vector_zeroth_sol-cond-all}
\begin{align}
 % & 0
 % = 2\, \delta\eta\, \mfh^\star_{vy}\, |_{\calH^+}
 % + 3\, i\, \delta\mfw\, \mfh^\star_{ry}\, |_{\calH^+}
%  ~,
%\label{eq:vector_zeroth_sol-cond-pre} \\
%   \Rightarrow \hspace{0.3truecm}
  & \frac{ \mfh^\star_{ry} }{ \mfh^\star_{vy} }\, \Bigg|_{r=1}
  = - \frac{2}{3}\frac{\deta}{i\dw}
%& & \left( \gamma := \frac{\delta\eta}{i\, \delta\mfw} \right)
  ~.
\label{eq:BC_shear}
\end{align}
%\end{subequations}
%

For the $p=2$ shear mode, one can obtain the special point solution. The solution is given by
\begin{subequations}
\label{eq:sol_shear}
\begin{align}
\mfh_{ry}^\star &= -C_1 + C_2
\left[
\frac{  (r^3+3r^2+2)
\exp\left\{ \sqrt{3}~\left( \arctan\frac{ 2r + 1 }{\sqrt{3}} - \frac{\pi}{2} \right)\right\} }
    {(r^2+r+1)^{3/2}} - 1
    \right]~, \\
\mfh_{vy}^\star &= (C_1+C_2) \frac{ 2r^3-3r^2-3r-2 }{ 2r } 
\nonumber \\
&- C_2 \frac{  (r+2)(2r^3-3r^2-2) 
\exp\left\{ \sqrt{3}~\left( \arctan\frac{ 2r + 1 }{\sqrt{3}} - \frac{\pi}{2} \right)\right\} }
{2r(r^2+r+1)^{1/2}}~.
%\label{eq:}
%
\end{align}
\end{subequations}
Asymptotically, $\mfh_{vy}^\star$ behaves as
\begin{align}
\frac{1}{r^2}\mfh_{vy}^\star \sim C_1 \left\{ 1 - \frac{3}{2r} - \frac{3}{2r^2} - \frac{ 1+\frac{21}{4}c^\star}{r^3} + \cdots \right\}~,
\quad(r\to\infty)
%\label{eq:}
%
\end{align}
where $c^\star:=C_2/C_1$. Imposing the boundary condition \eqref{eq:BC_shear}, one obtains 
\begin{align}
%
%\left. \frac{\mfh^\star_{ry}}{\mfh^\star_{vy}} \right|_{r=1} 
%&=
%\frac{ C_1+\frac{2}{9\sqrt{3}}C_2 }{ 3C_1+\frac{1}{2\sqrt{3}}C_2 } \\
\frac{C_2}{C_1} &= \frac{ 3(1+2\gamma) }{\sqrt{3}\,e^{-\frac{\pi}{2\sqrt{3}}} (2+3\gamma) - 3(1+2\gamma)}~.
%:= c_\text{shear}^\star
%\label{eq:}
%
\end{align}
The gauge-invariant variable $\mfh_{vy}$ is related to the master variable $Z_1$ by Kovtun and Starinets \cite{Kovtun:2005ev} as $\mfh_{vy}=r^2 Z_1$, so the Green's function depends on the ratio $c^\star$.

%%%%%%%%%
\section{Discussion}%\label{sec:}
%%%%%%%%%

%%%
\paragraph{Universality.}
%%%
We have shown that Green's functions of many systems are not unique at special points. In our examples, special points are located at $\nw_\star=-i$. From the bulk point of view, this universality comes from the fact that only the near-horizon behavior matters to the location. Thus, the existence of special points is not limited to asymptotically AdS black holes. The incoming mode is not uniquely defined at special points for those spacetimes as well. 

In our examples, $\nw_\star=-i$, but the sound mode has the special point at $\nw_\star=i$. It is interesting to study how the sound mode is different from these examples and to show under what conditions the location $\nw_\star=-i$ is universal. 

%Also, it is interesting to show the existence of special points from field theory point of view. The location $\nw_\star=\pm i$ obviously reminds of Matsubara frequencies of finite-temperature field theories. It is also interesting to explore the consequence of these special points. They are located in the lower-half $\omega$-plane in our examples, so they do not seem to indicate chaotic behaviors. 

% v1.3
Also, it is interesting to show the existence of special points from field theory point of view. The location $\nw_\star=\pm i$ obviously reminds of Matsubara frequencies of finite-temperature field theories. 

%It is also interesting to explore phenomenological implications of $\nw=-i$ special points. They are located in the lower-half $\omega$-plane, so they do not seem to indicate chaotic behaviors. 
%On the other hand, 
We did not find special points in the upper-half $\omega$-plane unlike the sound mode. But it does not necessarily imply that these modes do not show chaotic behaviors. It could imply that the pole-skipping fails to see chaotic behaviors for these modes. 
%But it does not mean that these modes do not show chaotic behaviors. One would expect chaotic behaviors for these modes as well. Rather, it probably means that the pole-skipping fails to see chaotic behaviors for these modes.  
So, the relation between the pole-skipping and chaotic behaviors is not robust and may be limited to the sound mode. 

%%%
\paragraph{Phenomenological implications.}
%%%
% v2
It is also interesting to explore phenomenological implications of $\nw_\star=-i$ special points. 
We do not have a complete physical interpretation, but we make a few brief remarks here. First of all, they are located in the lower-half $\omega$-plane, so they do not seem to indicate chaotic behaviors or exponentially growing modes. Instead, they should indicate decaying modes. 

%As an example, consider the Maxwell scalar mode. In this case, the special point is located in the lower-half $\omega$-plane with real $q_\star$. For a generic slope $\dq/\dw$, the special point is a pole. The late-time behavior of the perturbation is determined by the pole which is closest to the origin, and the perturbation decays as $e^{-t/\tau}$. For simplicity, let us assume the $\nw_\star=-i$ special point is such a pole. Then, $\tau\sim2\pi T$.

%However, by choosing the slope appropriately, the special point is no longer a pole. In this case, the late-time behavior is determined by the other poles, and the perturbation behaves differently. In principle, one could probe such a behavior.

%When $q$ is away from the special point, the Green's function has a pole with dispersion relation $\delta\omega \propto \delta q$. 
%Consider a fixed $q$ and approach the special point $q\to q_\star$. When $q$ is away from the special point, the Green's function has a pole with dispersion relation $\delta\omega \propto \delta q$. 

As an example, consider the Maxwell scalar mode. In this case, the special point is located in the lower-half $\omega$-plane with real $\nq_\star$. Let us follow the pole which becomes the special point. When $\delta\nq \neq 0$, the pole is located at $\nw=-i+\delta\nw$ and has the dispersion relation $\delta\nw \propto \delta \nq$.  
The late-time behavior of the perturbation is determined by the pole which is closest to the real axis, and the perturbation decays as $e^{-t/\tau}$ for a pole in the lower-half $\omega$-plane. For simplicity, let us assume the $\nw\sim-i$ pole is such a pole. Then, $\tau\sim2\pi T$.

As $\delta\nq\to0$, the pole approaches the special point along the dispersion relation. However, at the special point, the residue of the pole vanishes. Then, the perturbation no longer decays as $e^{-t/\tau}$. Instead, the late-time behavior is determined by the other poles, and the perturbation behaves differently. In principle, one could probe such a behavior by adding monochromatic perturbations with various $q$.

For the $\nw_\star=+i$ special point, $\nw_\star/\nq_\star$ is interpreted as the butterfly velocity. One could define $\nw_\star/\nq_\star$ in our examples as well, but it does not seem to represent physical propagations. First, this quantity is complex in general. Second, this ``velocity" can exceed the speed of light. In particular, $\nq_\star=0$ for the $p=2$ Maxwell field, so it diverges. 

%%%
\paragraph{Relation to hydrodynamic poles.}
%%%
For the sound mode, it is argued that the special point is related to a hydrodynamic pole as $\nw, \nq\to0$. Namely, write the sound mode Green's function as $G^R =b/a$. In hydrodynamic limit where $\omega, q \to 0$, $a(\omega,q)=0$ at the sound pole:
\begin{align}
\omega = c_s q -\frac{i}{2}\Gamma_s q^2 + \cdots~,
%\label{eq:}
%
\end{align}
where $c_s$ is the speed of sound and $\Gamma_s$ is the sound attenuation constant. Now, analytically continue the Green's function to a pure imaginary $q$ $(\Im q>0)$. The sound pole is now located in the upper-half $\omega$-plane. It is an unstable but is an unphysical pole. 
Then, follow the pole for finite $q$. At the special point, $b$ vanishes, the ``would-be" pole is absent, and the pole-skipping occurs. 

But not all systems have a hydrodynamic pole. A hydrodynamic pole typically arises as a consequence of conservation laws. In our examples, there are special points even for systems which lack a hydrodynamic pole (the bulk scalar field, the Maxwell vector, and the gravitational tensor mode). Thus, a special point is not necessarily related to a hydrodynamic pole. But it is interesting to study which quasinormal pole corresponds to the special point when $\nw,\nq\to0$.

%%%
\paragraph{Special points in 3 different senses.}
%%%
We used the word ``special point" in 3 different ways:
\begin{enumerate}
\item The point where the regular singularity at the horizon becomes a regular point.
\item The point where both near-horizon solutions become regular.
\item The point where the dual retarded Green's function are not unique. 
\end{enumerate}
They are related but are not equivalent. Needless to say, 3 is the original definition of the special point. In this paper, we use criterion 1 to locate special points. Criterion 1 is the sufficient condition of criterion 2 but is not the necessary condition. Namely, there may be the other points where both solutions become regular. Also, in our examples, 2 always leads to 3, but this is not necessarily true. It is desirable to find the necessary and sufficient condition of special points. 

%%%%%%%%%
%\section*{Acknowledgments}%\label{sec:}
%%%%%%%%%

%%%%%%%%%
\begin{acknowledgments}
This research was supported in part by a Grant-in-Aid for Scientific Research (17K05427) from the Ministry of Education, Culture, Sports, Science and Technology, Japan. 
\end{acknowledgments}
%%%%%%%%%

%\newpage
%\appendix 

\end{document}